\UseRawInputEncoding
\documentclass[final]{IEEEtran}
\usepackage{geometry}
\newgeometry{left=1.50cm,right=1.50cm,bottom=1.50cm,top=1.50cm}%
\usepackage{cite}
\usepackage{threeparttable}
\usepackage{graphicx}
\usepackage{picinpar}
\usepackage[cmex10]{amsmath}
\usepackage{amsmath,amsfonts,amssymb}
\usepackage{subfigure}
\usepackage{algorithm}
\usepackage{algorithmic}
\usepackage{stfloats}
\usepackage{bm}
\usepackage{amsthm}
\usepackage{color}

\begin{document}
\newtheorem{lemma}{Lemma}
\newtheorem{corol}{Corollary}
\newtheorem{theorem}{Theorem}
\newtheorem{proposition}{Proposition}
\newtheorem{problem}{Problem}
\newtheorem{definition}{Definition}
\newcommand{\e}{\begin{equation}}
\newcommand{\ee}{\end{equation}}
\newcommand{\eqn}{\begin{eqnarray}}
\newcommand{\eeqn}{\end{eqnarray}}

\title{CS-Based CSIT Estimation for Downlink Pilot Decontamination in Multi-Cell FDD Massive MIMO}%: Compressive Sensing
%\title{Novel Q-function approximation on SQNR formula for variance-matched scalar quantization of Gaussian source} %(CS)-Based Channel Estimation and Cooperative Precoding}

%

\author{Yikun Mei and Zhen Gao
\thanks{Y. Mei and Z. Gao are with School of Information and Electronics, Beijing Institute of Technology, Beijing 100081, China (E-mail: gaozhen010375@foxmail.com).}\\
 %\vspace*{-4.50mm}
}

\maketitle
\begin{abstract}
Efficient channel state information at transmitter (CSIT) for frequency division duplex (FDD) massive MIMO can facilitate its backward compatibility with existing FDD cellular networks. % is challenging, since each user has to estimate high-dimensional channel.
To date, several CSIT estimation schemes have been proposed for FDD single-cell massive MIMO systems, but they fail to consider inter-cell-interference (ICI) and suffer from downlink pilot contamination in multi-cell scenario. To solve this problem, this paper proposes a compressive sensing (CS)-based CSIT estimation scheme to combat ICI in FDD multi-cell massive MIMO systems. % from adjacent base stations (BSs).
% 这篇文章提出了基于压缩感知的信道估计方法，通过利用大规模多天线信道角度域成块稀疏特性
 Specifically, angle-domain massive MIMO channels exhibit the common sparsity over different subcarriers, and % due to the limited scatterers seen from the base station (BS) and the almost unchanged channel propagation property within the system bandwidth. %since multipath components between one user and the BS exhibit the small angle-domain spread seen from the BS, .
such sparsity is partially shared by adjacent users. % due to some common scatterers.
 By exploiting these sparsity properties,
 we design the pilot signal and the associated channel estimation algorithm under the framework of CS theory, where the channels associated with multiple adjacent BSs
 can be reliably estimated with low training overhead for downlink pilot decontamination.
 %Moreover, with the aid of previous CSIT,
 %the adaptive pilot design based closed-loop channel tracking closed is proposed for further improved channel estimation performance.
 %Moreover, partially sparse components are shared by users physically close to each other, which can leveraged for the further improved performance of CSIT acquisition.
 Simulation results verify the good downlink pilot decontamination performance of the proposed solution compared to its conventional counterparts in multi-cell FDD massive MIMO.
\end{abstract}
%\vspace*{-3mm}
\begin{IEEEkeywords}
 Frequency division duplex (FDD), massive MIMO, channel estimation, compressive sensing, pilot contamination.
 %\vspace*{-2mm}
\end{IEEEkeywords}

\IEEEpeerreviewmaketitle

%\vspace*{-6mm}
\section{Introduction}\label{S1}
%Massive multi-input multi-output (MIMO) employing hundreds
%of antennas at the base station (BS) has emerged
%as a promising technology to realize high-throughput green
%wireless communications \cite{MMIMOover}.
%The exponential increase of mobile data traffic is challenging current
%cellular networks, including the most
%advanced fourth generation (4G) network.
%It has
%been the consensus that future 5G networks
%should achieve the 1000-fold increase in system
%capacity~\cite{myWC,key}. To realize such an aggressive 5G version,
%Massive MIMO can substantially boost the spectrum and energy efficiency by
%exploiting hundreds of degrees of spatial freedom~\cite{Massve_MIMO_101}. %Hence, massive MIMO
%has been considered as a promising key technique for 5G cellular network \cite{ }.
%In massive MIMO,
Reliable channel state information at transmitter (CSIT)
is essential to fully exploit potential advantages of massive MIMO. For
time division duplex (TDD) massive MIMO, CSIT can be acquired in the uplink
by leveraging the channel reciprocity, where the channels of dozens of users can
be easily acquired at base station (BS) with hundreds of antennas~\cite{Massve_MIMO_101,GFF_CM}.
However, CSIT for frequency division duplex (FDD) massive MIMO can be more challenging, since
single-antenna users have to acquire and feedback the high-dimensional channels to the BS~\cite{MXS, FJ_TVT,WS_TVT,Shen,HY_Tcom,Rao1,Rao2,WS_CL,my_TSP}.

To date, there have been several CSIT estimation schemes
proposed for FDD massive MIMO to facilitate its backward compatibility
with current cellular networks dominated by FDD~\cite{FJ_TVT,WS_TVT,Shen,HY_Tcom,Rao1,Rao2,WS_CL,my_TSP}.
%In   communication systems, one radio frequency (RF) chain required by each antenna can be impossible, since the hardware cost and power consumption of RF chains in   can be prohibitive. While conventional   multi-antenna systems with single RF chain cannot realize spatial multiplexing. To solve this issue, the   massive MIMO with much smaller number of RF chains but still large number of antennas has been emerging for its feasibility, and the analog phase shifter network can be used to realize spatial multiplexing in RF.
Specifically, \cite{FJ_TVT} proposed an estimated covariance-assisted minimum mean square error (MMSE) estimator for channel estimation of FDD massive MIMO, but it may be inaccurate to obtain downlink covariance matrix from uplink channel information. \cite{WS_TVT,Shen,WS_CL} proposed the compressive sensing (CS)-based downlink channel estimation by assuming the delay-domain sparsity of massive MIMO channels, but such assumption may not hold in indoor scenarios due to rich scatterers at the user side. \cite{HY_Tcom,Rao1,Rao2,my_TSP} proposed the CS-based CSIT estimation schemes by assuming the sparsity of angle-domain massive MIMO channels. However, \cite{HY_Tcom,Rao1,Rao2} are limited to narrow-band systems without considering practical broad-band systems, while \cite{Rao2,my_TSP} only consider the signal-user CSIT estimation and fail to exploit the channel correlation of multiple adjacent users. %But this scheme also cannot combat ICI in multi-cell scenario.
By exploiting the channel sparsity in both angle and delay domains,  an efficient CSIT estimation scheme taking the spatial-wideband effect of massive MIMO system into account is proposed in \cite{GFF_TSP}.
Furthermore, existing schemes~\cite{FJ_TVT,WS_TVT,Shen,HY_Tcom,Rao1,Rao2,WS_CL,my_TSP,GFF_TSP} only consider the single-cell scenario, and they may suffer from downlink pilot contamination due to inter-cell-interfere (ICI).

In this paper, we consider the practical multi-cell FDD massive MIMO systems.
In such scenario, users in target cell will receive the downlink pilot from adjacent cells, which will contaminate the downlink channel estimation of the target cell and thus degrade the system performance.
This phenomenon is termed as the \emph{downlink pilot contamination of multi-cell FDD massive MIMO}, while
conventional CSI acquisition schemes either only consider ICI in TDD massive MIMO or fail to consider ICI in FDD massive MIMO.
To this end, we propose a CS-based
CSIT estimation scheme to alleviate the pilot contamination in multi-cell FDD massive MIMO systems.
Particularly,
we observe that angle-domain massive MIMO channels exhibit the common sparsity over different
subcarriers due to the limited number of scatterers seen from the BS and the very similar scatterers experienced
by different subcarriers. Moreover, such sparsity is
partially shared by adjacent users due to some common scatterers. By jointly
exploiting these sparsity properties of massive MIMO channels in the angular domain,
under the framework of CS theory, we design the pilot signal and CS-based channel estimator for multi-cell
FDD massive MIMO.
The proposed scheme can reliably acquire the channels associated with multiple adjacent
BSs with low training overhead for downlink pilot decontamination. Simulation results
confirm that the proposed solution outperforms existing schemes
in multi-cell FDD massive MIMO systems.

Notation: the boldface lower and upper-case symbols denote column vectors and matrices, respectively. The Moore-Penrose inversion, transpose,
 and conjugate transpose operators are given by $(\cdot )^{\dag}$, $(\cdot )^{\rm T}$ and
 $(\cdot )^{*}$, respectively. %$(\cdot )^{-1}$ is the inverse operator.
 %The $\ell_{0}$-norm and $\ell_{2}$-norm are
 %given by $\|\cdot\|_0$ and $\|\cdot\|_2$, respectively, and
 $\left| \Gamma \right|_c$ is the
 cardinality of the set $\Gamma$. %The support set of the vector $\mathbf{a}$ is denoted by
% ${\rm supp}\{\mathbf{a}\}$. %The rank of ${\bf A}$ is denoted by ${\rm rank}\{{\bf A}\}$, while
 ${\rm E}\{\cdot \}$ is the
 expectation operator.
$\left( {\bf a} \right)_{\Gamma}$ denotes the entries of $\mathbf{a}$ whose
 indices are defined by $\Gamma$, while $\left( {\bf A}\right)_{:,k}$ denotes the $k$th column of the matrix $\mathbf{A}$. $\left[ {\bf a} \right]_i$ denotes the $i$th entry of the
 vector $\mathbf{a}$, and $\left[ {\bf A} \right]_{i,j}$ denotes the $i$th-row and
 $j$th-column element of the matrix $\mathbf{A}$. Finally, ${\Omega ^c}$ is the complementary set of $\Omega $.

 %\vspace*{-2mm}
\section{System Model}\label{S2}

We consider a multi-cell FDD massive MIMO system composed of $L$ hexagonal cells,
and each cell consists of a central $M$-antenna BS and $N$ single-antenna users with $N \ll M $ \cite{Massve_MIMO_101}.
%In the downlink channel estimation,
For the $k$th user in the $\tilde l$th cell, the received downlink signal of the $p$th subcarrier can be expressed as
  %\vspace*{-1mm}
\begin{equation}\label{equ:H} % eq1
\begin{small}
y_{k,\tilde l,p}^{}\!\! =\!\! {\bf{x}}_{\tilde l,p}^{\rm{T}}{\bf{h}}_{k,\tilde l,p}^{} + \sum\nolimits_{l = 0,l \ne \tilde l}^{L-1} {} {\bf{x}}_{l,p}^{\rm{T}}{\bf{h}}_{k,l,p}^{} + {v_{k,\tilde l,p}},1 \le p \le P,
\end{small}  %\vspace*{-1mm}
\end{equation}
 where ${\bf h}_{k,l,p}\in \mathbb{C}^{M\times 1}$ denotes the downlink channel of the $p$th subcarrier between
 the $k$th user and the $l$th BS, ${\bf x}_{l,p} \in \mathbb{C}^{M \times 1}$
 is the transmitted signal from the $l$th BS, $v_{k,\tilde l,p}$ is additive
 white Gaussian noise (AWGN), and $P$ is the size of one OFDM symbol. From (\ref{equ:H}), it can be observed that the reliable estimation of ${\bf{h}}_{k,\tilde l,p}$ is challenging due to two following reasons. First, the estimation of $M$-dimensional ${\bf{h}}_{k,\tilde l,p}$ will lead to the prohibitively high training overhead. Second, the $k$th user of the $\tilde l$th cell suffers from ICI, i.e., $\sum\nolimits_{l = 0,l \ne \tilde l}^{L-1} {} {\bf{x}}_{l,p}^{\rm{T}}{\bf{h}}_{k,l,p}^{}$.

\begin{figure*}[htbp]
\centering
\includegraphics[width=10cm]{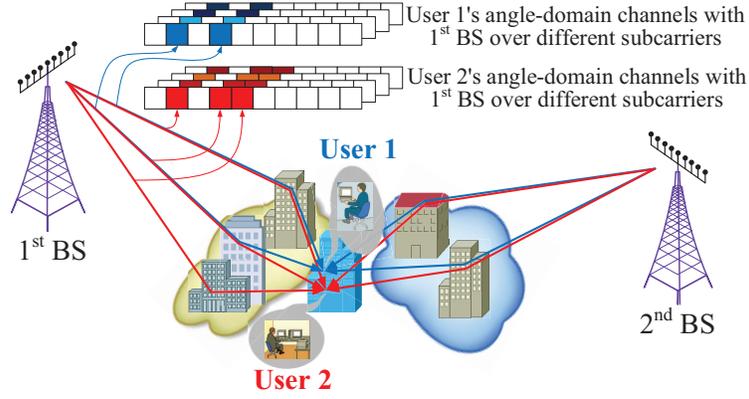}
%\vspace*{-3mm}
\caption{Illustration of the angle-domain sparsity of massive MIMO channels.} \label{fig:FDD_LS_MIMO}
%\vspace*{-6mm}
\end{figure*}

For massive MIMO systems as shown in Fig. \ref{fig:FDD_LS_MIMO}, the BS is usually elevated high with few scatterers around, while users are located
 at low elevation with relatively rich local scatterers, which leads the multipath components of the channels associated with
 one user to concentrate on the limited angle seen from the BS side \cite{Rao2,my_TSP}.
 Based on this phenomenon, \cite{Rao2,my_TSP} assume that the angle-domain massive MIMO channel vectors ${\bf{\tilde h}}_{k,l,p}^{} = {{\bf{F}}}^*{\bf{ h}}_{k,l,p}^{}$ exhibit the {sparsity}, where ${\bf F}\in \mathbb{C}^{M\times M}$ is the unitary matrix representing the
 transformation matrix of the angular domain at the BS side. Such sparsity indicates that only a small part of elements of the angle-domain channel vector ${\bf \tilde h}_{k,l,p}$ contain almost all the multipath components between the $l$th BS and the $k$th user, i.e., ${\left| {{\Omega _{k,l,p}}} \right|_c} \ll M$, where%\vspace*{-2mm}
   %with ${\left| {{\Omega _{k,l,p}}} \right|_c} \ll M$, where ${\bf{\tilde h}}_{k,l,p}^{} = {{\bf{F}}}^*{\bf{ h}}_{k,l,p}^{}$, ${\bf F}\in \mathbb{C}^{M\times M}$ is the unitary matrix representing the
% transformation matrix of the angular domain at the BS side,
    \begin{equation}\label{equ:support}
{\Omega _{k,l,p}}{\rm{ = supp}}\{ {{\bf{\tilde h}}_{k,l,p}}\} \! =\! \left\{\! {m\!:\!\!{{\left\|\!{{{[ {{{\bf{\tilde h}}_{k,l,p}}} ]}_m}} \right\|}_2}\!\!\!\! > \!\!{p_{{\rm{th}}}},\! 1 \le \!\! m \!\! \le\!\! M} \! \right\}\!\!,%\vspace*{-2mm}
\end{equation}
and $p_{\rm th}$ is a threshold according to AWGN \cite{Shen}. Moreover, since channels of different subcarriers experience the very similar scatterers, they share the same sparsity pattern~\cite{my_TSP}, i.e.,%\vspace*{-2mm}
 \begin{equation}\label{equ:set2}
{\Omega _{k,l,1}} = {\Omega _{k,l,2}} =  \cdots {\rm{ = }}{\Omega _{k,l,P}} = {\Omega _{k,l}}.%\vspace*{-2mm}
 \end{equation}

 Additionally, for a group of $K$ users physically close to each other as illustrated in Fig. \ref{fig:FDD_LS_MIMO}, their angle-domain channels share
 the partially common sparsity \cite{Rao1}, which can be expressed as%\vspace*{-2mm}
 \begin{equation}\label{equ:set}
 \mathop  \cap \limits_{k = 1}^K {\Omega _{k,l}} = {\Omega _c} \ne \phi.%\vspace*{-2mm}
 \end{equation}
It should be pointed out that $N$ users served by the BS using the same time-frequency resource usually come from different user groups for the improved performance \cite{Massve_MIMO_101}.
%\vspace*{-2mm}

%\vspace*{-2mm}
\section{Proposed CS-Based CSIT Estimation Scheme}\label{S3}% for Downlink Pilot Decontamination
%This section proposes a CS-based CSIT estimation scheme, whereby the angle-domain sparsity of massive MIMO channels
%is exploited to jointly acquire the channels of multiple adjacent BSs with low training overhead for pilot decontamination.
The proposed scheme includes the CS-based design of downlink multi-cell pilot and channel estimation algorithm, and both of them are significant for downlink pilot decontamination.
By leveraging the angle-domain sparsity of massive MIMO channels, the proposed scheme can jointly acquire the channels of multiple adjacent BSs with low training overhead, which can mitigate the downlink pilot contamination.

%\vspace*{-3mm}
\subsection{Pilot Training for CSIT Estimation in Multi-Cell Scenario}\label{S3.1}
In the proposed scheme, each BS transmits the off-line designed downlink pilot for CSIT estimation, and the received downlink pilot signal at users can be fed back to their respective BSs via the uplink feedback channels. Here the uplink feedback channels are assumed to be AWGN channels after the uplink channel estimation and equalization \cite{Rao1}. For the $k$th user of the central target cell ($l=0$) in the $t$th time slot, the received pilot signal fed back to the BS can be expressed as%%\vspace*{-1.5mm}
  \begin{equation}\label{equ:channelmode1} % eq3
\!\!\!\!\!\!  \begin{small}
\begin{array}{l}
{r_{k,p}^t} = \sum\limits_{l = 0}^{L - 1} {({{\bf{s}}_{l,p}^t})^{\rm{T}}{\bf{h}}_{k,l,p}^{}}  + {w_{k,p}^t}= \sum\limits_{l = 0}^{L - 1}\! {({\bf{s}}_{l,p}^t)^{\rm{T}}{\bf{h}}_{k,l,p}^{}\delta \left( {{\rho _{k,l}} \!> \!{\rho _{{\rm{th}}}}} \right)} \!
\\ ~~~~~~~~~~~~~~+\!\!\! \sum\limits_{l = 0}^{L - 1}\! {({\bf{s}}_{l,p}^t)^{\rm{T}}{\bf{h}}_{k,l,p}^{}} \delta \left( {{\rho _{k,l}} \!\le\! {\rho _{{\rm{th}}}}} \right)\! +\! {w_{k,p}^t}\\
~~~~~= \sum\limits_{l = 0}^{L - 1} {({\bf{s}}_{l,p}^t)^{\rm{T}}{\bf{h}}_{k,l,p}^{}\delta \left( {{\rho _{k,l}} > {\rho _{{\rm{th}}}}} \right)} + {{\tilde w}_{k,p}^t},
\end{array}
\end{small}%%\vspace*{-1.5mm}
\end{equation}
%where superscript $p$ and subscript $t$ denote indices of subcarrier and time slot, respectively,
where $\delta \left( {\cdot} \right)$ is Dirac delta function, ${\bf{s}}_{l,p}^t$ is the downlink pilot of the $l$th cell in the $t$th time slot, $\rho_{\rm th}$ is a predefined signal-to-noise-ratio (SNR) threshold, $\rho_{k,l}$ is the $k$th user's SNR associated with the $l$th BS, ${w_{k,p}^t}$ is the effective noise including the downlink channel and uplink feedback channel \cite{Rao1}, and ${{\tilde w}_{k,p}^t} = \sum\nolimits_{l = 0}^{L - 1} {({\bf{s}}_{l,p}^t)^{\rm{T}}{\bf{h}}_{k,l,p}^{}} \delta \left( {{\rho _{k,l} } \le {\rho _{{\rm{th}}}}} \right) + {w_{k,p}^t}$. %We consider the $k$th user will   estimate the channels

Due to the angle-domain sparsity of massive MIMO channel vectors as discussed in Section \ref{S2}, (\ref{equ:channelmode1}) can be rewritten as
  \begin{equation}\label{equ:channelmode2} % eq3
\begin{array}{l}
r_{k,p}^t = \sum\limits_{l \in \Pi_k }^{} {{{({\bf{s}}_{l,p}^t)}^{\rm{T}}}{\bf{F\tilde h}}_{k,l,p}^{}}  + {{\tilde w}_{k,p}^t}= \sum\limits_{l \in \Pi_k }^{} {{\bm{\phi }}_{l,p}^t{\bold{\tilde h}}_{k,l,p}^{}}  + {{\tilde w}_{k,p}^t}\\
~~~~ = {{\bm{\theta }}_{k,p}^t}{\bf{\bar {\tilde h}}}_{k,p}^{} + \tilde w_{k,p}^t,
\end{array}%%\vspace*{-3mm}
\end{equation}
where
 \begin{equation}\label{equ:channelmode3} % eq3
\left\{ \begin{array}{l}
 \Pi_k  = \left\{ {l:{\rho _{k,l}} > {\rho _{{\rm{th}}}},0 \le l \le L-1} \right\},\\
 {{\bm{\phi }}^t_{l,p}} = {({\bf{s}}_{l,p}^t)^{\rm{T}}}{\bf{F}} \in \mathbb{C}^{1 \times M},\\
{{\bm{\theta }}_{k,p}^t} = [ {{\bm{\phi}} _{\Pi_k \left( 1 \right),p}^t,{\bm{\phi}} _{\Pi_k  \left( 2 \right),p}^t, \cdots ,{\bm{\phi}} _{\Pi_k  ( {{{\left| \Pi_k   \right|}_c}} ),p}^t} ]\in \mathbb{C}^{1 \times {M{\left| \Pi_k   \right|}_c}},\\
{\bf{\bar{ \tilde h}}}_{k,p}^{} = {[ {{\bf{\tilde h}}_{k,\Pi_k  \left(\! 1 \!\right),p}^{\rm{T}}\!,\!{\bf{\tilde h}}_{k,\Pi_k  \left( \!2 \!\right),p}^{\rm{T}}, \!\cdots\!,\!{\bf{\tilde h}}_{k,\Pi_k  ( {{{\left| \Pi_k   \right|}_c}}) ,p}^{\rm{T}}} ]^{\rm{T}}}\!\!\!\in\!\! \mathbb{C}^{M{{\left| \Pi_k   \right|}_c} \!\times \! 1},
\end{array} \right.
\end{equation}
$\Pi_k  \left( i \right)$ denotes the $i$th element of the set $\Pi_k $, which can be acquired by comparing the received SNRs associated with different BSs and $\rho_{\rm th}$ at the $k$th user, and then fed them back to BSs for the following CSIT estimation.

Moreover, due to the temporal channel correlation,
the channel ${\bf{h}}_{k,l,p}^{}$ is considered to be unchanged
in $G$ successive OFDM symbols within the channel coherence time \cite{Shen}. By jointly collecting the feedback pilots in $G$ successive OFDM symbols, we can obtain the aggregate feedback signal%%%\vspace*{-2mm}
\begin{align}\label{equ:joint_process2} % eq 9
{\bf{ r}}_{k,p}^{[G]}  = {{\bm{\Theta }}_{k,p} ^{[G]}}{\bf{\bar {\tilde h}}}_{k,p}^{} +  {\bf \tilde w}_{k,p}^{[G]},
%%\vspace*{-1.5mm}
\end{align}
where we have
  \begin{equation}\label{equ:joint_process3}
\left\{ \begin{array}{l}
{\bf{r}}_{k,p}^{[G]} = {[{({{r}}_{k,p}^1)^{\rm{T}}},{({{r}}_{k,p}^2)^{\rm{T}}}, \cdots ,{({{r}}_{k,p}^G)^{\rm{T}}}]^{\rm{T}}} \in {\mathbb{C}^{G \times 1}},\\
{\bf{\Theta}}_{k,p}^{[G]} = {[{({\bm{\theta }}_{k,p}^{1})^{\rm{T}}},{({\bm{\theta }}_{k,p}^{2})^{\rm{T}}}, \cdots ,{({\bm{\theta }}_{k,p}^{G})^{\rm{T}}}]^{\rm{T}}} \in {\mathbb{C}^{G\times M{{\left| \Pi_k   \right|}_c}}},\\
{\bf{\tilde w}}_{k,p}^{[G]} = {[\tilde w_{k,p}^1,\tilde w_{k,p}^2, \cdots ,\tilde w_{k,p}^G]^{\rm{T}}} \in {\mathbb{C}^{G \times 1}}.
\end{array} \right.%\vspace*{-1.5mm}
\end{equation}
 % and the system's SNR
 %can be defined as $\mbox{SNR}={\rm E}\Big\{\Big\| {{\bm{\Theta }}_{k,p} ^{[G]}}{\bf{\bar {\tilde h}}}_{k,p}^{}\Big\|_2^2\Big\} \Big/ {\rm E}\Big\{\Big\| {\bf \tilde w}_{k,p}^{[G]}\Big\|_2^2\Big\}$ according to (\ref{equ:joint_process2}).
%\vspace*{-5mm}
\subsection{CS-Based CSIT Estimation Algorithm}\label{S3.2}
To reliably acquire the channel vector ${\bf{\bar {\tilde h}}}_{k,p}^{}$ from (\ref{equ:joint_process2}), the training overhead $G$ required by conventional algorithms,
 e.g., the least squares (LS) algorithm, is usually proportional to $M{{\left| \Pi_k   \right|}_c}$, the
 dimension of ${\bf{\bar {\tilde h}}}_{k,p}^{}$. Usually, $G\ge M{{\left| \Pi_k   \right|}_c} $ is required,
 which leads $G$ to be much larger than the channel coherence time, and otherwise results in the
 poor channel estimation performance~\cite{FJ_TVT}.

 Fortunately, the angle-domain sparsity of massive MIMO channel ${\bf{\tilde h}}_{k,l,p}^{}$ implies that the aggregate angle-domain channel ${\bf{ {\bar {\tilde h}}}}_{k,p}^{}$ also has the sparsity according to (\ref{equ:channelmode3}), which
 motivates us to leverage the CS theory to estimate high-dimensional ${\bf{ {\bar {\tilde h}}}}_{k,p}^{}$ from low-dimensional ${\bf{ r}}_{k,p}^{[G]}$ in (\ref{equ:joint_process2}). Moreover, the common sparsity shared by $\{{\bf{ {\bar {\tilde h}}}}_{k,p}^{}\}_{p=1}^P$ for the $k$th user and the partially common sparsity shared by $\{{\bf{ {\bar {\tilde h}}}}_{k,p}^{}\}_{k=1}^K$ for $K$ users in the same group can be leveraged for the further improved performance.
Specifically, we consider the partially common support shared by the
$K$ users physically close to each other, i.e.,%\vspace*{-1.5mm}
  \begin{equation}\label{equ:joint_process4}
  {\bf{ R}}_p^{[G]}  = {{\bm{\Theta }} _p^{[G]}}{\bf{\bar {\tilde H}}}_p^{} +  {\bf \tilde W}_p^{[G]}, 1\le p \le P,
  %\vspace*{-1.5mm}
\end{equation}
where we have
  \begin{equation}\label{equ:joint_process5}%\vspace*{-1.5mm}
\left\{ \begin{array}{l}
{\bf{R}}_{p}^{[G]} = \left[ {{\bf{r}}_{1,p}^{[G]},{\bf{r}}_{2,p}^{[G]}, \cdots ,{\bf{r}}_{K,p}^{[G]}} \right] \in {\mathbb{C}^{G \times K}},\\
{\Pi _1} = {\Pi _2} =  \cdots  = {\Pi _K} = \Pi, \\
{\bm \Theta}_{1,p}^{[G]} = {\bm \Theta} _{2,p}^{[G]} =  \cdots  = {\bm \Theta} _{K,p}^{[G]} = {\bm \Theta} _{p}^{[G]} \in {\mathbb{C}^{G\times {M{\left| \Pi   \right|}_c}}},\\
{\bf{\bar {\tilde H}}}_p = \left[ {{{{\bf{\bar {\tilde h}}}}_{1,p}},{{{\bf{\bar {\tilde h}}}}_{2,p}}, \cdots ,{{{\bf{\bar {\tilde h}}}}_{K,p}}} \right]\in {\mathbb{C}^{ M{{\left| \Pi   \right|}_c}\times K}},\\
{\bf{\tilde W}}_p^{[G]} = {\left[{\bf{\tilde w}}_{1,p}^{[G]},{\bf{\tilde w}}_{2,p}^{[G]}, \cdots ,{\bf{\tilde w}}_{K,p}^{[G]}\right]^{\rm{T}}} \in {\mathbb{C}^{G \times K}}.
\end{array} \right.%%\vspace*{-1.5mm}
\end{equation}
Note that since $K$ users in the same group are physically close to each other and their received signals from the same BS experience very similar large-scale fading, we can approximately obtain ${\rho _{l,1}} = {\rho _{l,2}} =  \cdots  = {\rho _{l,K}}$, and thus the second and third equations in (\ref{equ:joint_process5}) hold.

Given the measurements (\ref{equ:joint_process4}) and the sparse constraints (\ref{equ:set2}) and (\ref{equ:set}), the CSI matrix $\{{\bf{\bar {\tilde H}}}_p^{}\}_{p=1}^P$ can be acquired by
solving the following optimization problem%\vspace*{-1.5mm}
\begin{align}\label{equ:target_func} % eq 10
\begin{small}
\!\!\!\!\!\!\!\begin{array}{*{20}{l}}
{\mathop {\min }\limits_{{\bf{\bar {\tilde H}}}_p^{},1 \le p \le P} \!\!\!\!\sum\nolimits_{p = 1}^P {{{\left\| {{\bf{\bar{ \tilde H}}}_p^{}} \right\|}_{0,2}}} {\rm{\!= }}\mathop {\min }\limits_{{\bf{\bar {\tilde H}}}_p^{},1 \le p \le P}\!\!\!\!\!\! \sum\nolimits_{p = 1}^P\!\! {{{\left( {\sum\nolimits_{k = 1}^K \!\!{\left\| {{\bf{\bar {\tilde h}}}_{k,p}^{}} \right\|_0^2} } \right)}^{1/2}}} }\\
~~~~{{\rm{s}}.{\rm{t}}.~{\bf{R}}_p^{[G]} = {\bm{\Theta}} _p^{[G]}{\bf{\bar {\tilde H}}}_p^{},{\Omega _{k,l,p}} = {\Omega _{k,l}},\forall p,\mathop  \cap \limits_{k = 1}^K {\Omega _{k,l}}  \ne \phi}.
\end{array}
\end{small}%\vspace*{-1.5mm}
\end{align}

To solve the optimization problem (\ref{equ:target_func}), developed from the classical CS algorithm orthogonal matching pursuit (OMP), as shown in \textbf{Algorithm 1}, we propose a joint multi-user multi-carrier orthogonal matching pursuit (J-MUMC-OMP) algorithm.
Specifically, lines 1-3 initialize the variables;
lines 6 identifies the most possible angle-domain element by
leveraging the sparsity constraints (\ref{equ:set2}) and (\ref{equ:set});
lines 7-8 estimate the elements according to updated support set; lines 9-10 imply that if all $K$ users' $\rho$th angle-domain elements are dominated by AWGN, the iteration stops since the channel sparsity level is over-estimated; while in lines 11-12, for users whose $\rho$th angle-domain elements are dominated by AWGN, we delete the index $\rho$ and re-estimate the associated elements; lines 14-15 update the residue; line 16 indicates that if the residue
of the current iteration is larger than that of the last
iteration, stopping the iteration can help the algorithm to acquire the good mean square error (MSE) performance.

The proposed J-MUMC-OMP algorithm has several distinctive features as follows.
First, the proposed J-MUMC-OMP algorithm can jointly estimate the sparse signals $\{{\bf{\bar {\tilde h}}}_{k,p}^{}\}_{p=1,k=1}^{P,K}$, $\forall k$
by exploiting their common sparsity over different subcarriers. %ally common sparsity.
Second, the partially common sparsity of $K$ users' sparse channels $\{{\bf{\bar {\tilde h}}}_{k,p}^{}\}_{k=1}^{K}$, $\forall p$ is also considered for the
further improved performance.
Third, we provide the stopping criteria to adaptively acquire the sparsity level of channels.
By contrast, the classical
orthogonal matching pursuit (OMP) algorithm requires
the sparsity level without considering these sparsity properties, while the
joint-OMP algorithm proposed in \cite{Rao1} fails to leverage the common sparsity over
 different subcarriers.

\begin{algorithm}[tp]
\begin{small}
%\algsetup{linenosize=\scriptsize}\scriptsize
\renewcommand{\algorithmicrequire}{\textbf{Input:}}
\renewcommand\algorithmicensure {\textbf{Output:} }
\caption{Proposed J-MUMC-OMP Algorithm.}
\label{alg:Framwork} % Alg 1
\begin{algorithmic}[1]
\REQUIRE
Noisy measurement matrix $  {\bf{ R}}_p^{[G]}$, sensing matrix ${{\bm{\Theta }}_p ^{[G]}}$, $\forall p$, and the termination threshold ${\gamma _{{\rm{th}}}}$.% in (\ref{equ:compact2}).%, common sparsity level $K$, maximum channel length $L$, number of transmit antenna $M$, adjacent OFDM symbol $R$.
\ENSURE
The estimation of channel matrix ${\bf{\bar {\tilde H}}}_p^{}$, $\forall p$. \\
%\Inization ~~\\                          %算法的输入参数：Initialization
\STATE $i=0$;~$\{$Initialize the iteration index $i$$\}$
\STATE $\left\{ {{\Omega _k^{i}}} \right\}_{k = 1}^K = \phi $;~$\{$Initialize the support sets of $K$ users' aggregate channel vectors$\}$
\STATE ${\bf{Z}}_p^{i}  =  {\bf{ R}}_p^{[G]}$;~$\{$Initialize the residue$\}$
%\REPEAT
%\UNTIL{$\sum\nolimits_{p=1}^{P} \left\| \left[ {\bf c}_p \right]_{l_{\min}} \right\|_2^2 / {P}
% < p_{\rm{th}}$}
\REPEAT
\STATE $i=i+1$;
\STATE $\rho  = \arg \mathop {\max }\limits_{\tilde \rho } \left\{ {\sum\nolimits_{p = 1}^P {\sum\nolimits_{k = 1}^K {\left\| {{{\left[ {{{\left( {{\bm{\Theta}} _p^{[G]}} \right)}^*}{{\left[ {{\bf{Z}}_p^{i - 1}} \right]}_{:,k}}} \right]}_{\tilde \rho }}} \right\|_2^2} } } \right\}$;
\STATE ${\Omega _k^i} = {\Omega _k^{i-1}} \cup \rho ,\forall k$;
\STATE ${({{\bf{g}}_{k,p}})_{\Omega _k^i}} = ({\bm{\Theta}} _p^{[G]})_{\Omega _k^i}^\dag {\left[ {{\bf{R}}_p^{[G]}} \right]_{:,k}}$, ${({{\bf{g}}_{k,p}})_{{{(\Omega _k^i)}^c}}} = {\bf{0}}$, $\forall k,p$;
\IF {$\sum\nolimits_{p = 1}^P {\left\| {{{\left[ {{{\bf{g}}_{k,p}}} \right]}_\rho }} \right\|} _2^2/P < {\gamma _{{\rm{th}}}}$, $\forall k$}
    \STATE Quit iteration;
\ELSIF {there exists $k$ meeting $\sum\nolimits_{p = 1}^P {\left\| {{{\left[ {{{\bf{g}}_{k,p}}} \right]}_\rho }} \right\|} _2^2/P < {\gamma _{{\rm{th}}}}$}
\STATE ${\Omega _k^i}\! =\! {\Omega _k^{i-1}}$, ${({{\bf{g}}_{k,p}})_{\Omega _k^i}} \! \! =\! ({\bm{\Theta}} _p^{[G]})_{\Omega _k^i}^\dag {\left[ {{\bf{R}}_p^{[G]}} \right]_{:,k}}$, ${({{\bf{g}}_{k,p}})_{{{(\Omega _k^i)}^c}}}\! \! =\! \! {\bf{0}}$, $\forall p$; for $k$ satisfy the above condition;
\ENDIF
\STATE ${\bf{G}}_p^i = \left[ {{{\bf{g}}_{1,p}},{{\bf{g}}_{2,p}}, \cdots ,{{\bf{g}}_{K,p}}} \right]$, $\forall p$;
\STATE ${{\bf{Z}}_p^i} = {{\bf{R}}_p^{[G]}} - {{\bm \Theta} _p^{[G]}}{\bf{G}}_p^i$, $\forall p$;
%\STATE {\textbf{if}} ${\left\| {{{\bf{Z}}^i}} \right\|_F} > {\left\| {{{\bf{Z}}^{i - 1}}} \right\|_F}$
\UNTIL{$\sum\nolimits_{p = 1}^P {{{\left\| {{\bf{Z}}_p^i} \right\|}_F}}  \ge \sum\nolimits_{p = 1}^P {{{\left\| {{\bf{Z}}_p^{i - 1}} \right\|}_F}} $};
\STATE ${\bf{\bar {\tilde H}}}_p^{}={\bf{G}}_p^{i-1}$, $\forall p$;
\end{algorithmic}
\end{small}
\end{algorithm}

%\vspace*{-2mm}
\subsection{CS-Based Downlink Pilot Design in Multi-Cell Scenario}\label{S3.2}
The design of measurement matrices ${{\bm{\Theta }} _p^{[G]}}$ for different $p$'s in
 (\ref{equ:target_func}) are important to ensure the reliable channel estimation in CS theory.
Owing to ${\bf{\Theta}}_{p}^{[G]} = {[{({\bm{\theta }}_{p}^{1})^{\rm{T}}},{({\bm{\theta }}_{p}^{2})^{\rm{T}}}, \cdots ,{({\bm{\theta }}_{p}^{G})^{\rm{T}}}]^{\rm{T}}}$, ${{\bm{\theta }}_{p}^t} = [ {{\bm{\phi}} _{\Pi \left( 1 \right),p}^t,{\bm{\phi}} _{\Pi  \left( 2 \right),p}^t, \cdots ,{\bm{\phi}} _{\Pi  ( {{{\left| \Pi   \right|}_c}} ),p}^t} ]$, and  ${{\bm{\phi }}^t_{l,p}} = {({\bf{s}}_{l,p}^t)^{\rm{T}}}{\bf{F}}$, we observe that ${{\bm{\Theta }} _p^{[G]}}$, $\forall p$ are only
   determined by the pilot signals $\{ {{\bf{s}}_{l,p}^t} \}_{l = 0,p = 1,t = 1}^{L-1,P,G}$.

 According to \cite{STR_CS}, a measurement matrix whose elements follow an independent identically distributed (i.i.d.)
 Gaussian distribution can achieve the good performance for sparse signal recovery. Furthermore, diversifying measurement matrices
 ${{\bm{\Theta }} _p^{[G]}}$, $\forall p$ can further improve the recovery performance of sparse signals when multiple sparse signals with the (partially) common sparsity are jointly recovered \cite{STR_CS}. Specifically, we consider
 each element of pilot signals can be off-line designed as
\begin{align}\label{equ:pilot_element1} % eq 20
\!\!\!\!\!\!{\left[ {{\bf{s}}_{l,p}^t} \right]_m}\!\!\!\!\! = \!{e^{{\rm{j}}{\theta _{m,l,p,t}}}}\!\!,{\mkern 1mu} \!1 \!\!\le m \!\le \!\!M,\!1 \!\le \!t \!\le \!G\!,\!1\! \le\! p \!\le \!P,\!0 \!\le \!l \!\le \!L\!-\!1,
\end{align}
 where ${\theta _{m,l,p,t}}$ follows the
 i.i.d. uniform distribution in $[0, ~ 2\pi)$. It is straightforward to prove that
 the designed pilot signals (\ref{equ:pilot_element1}) can ensure elements of
 ${{\bm{\Theta }} _p^{[G]}}$, $\forall p$, to obey the i.i.d. complex
 Gaussian distribution with zero mean and unit variance.
 Hence, the proposed pilot signal design is optimal for the reliable compression
 and recovery of sparse angle-domain channels under the framework of CS theory.
 %\vspace*{-3mm}

 \begin{figure}[tp!]
\vspace*{4mm}
\begin{center}
\includegraphics[width=1\columnwidth, keepaspectratio]{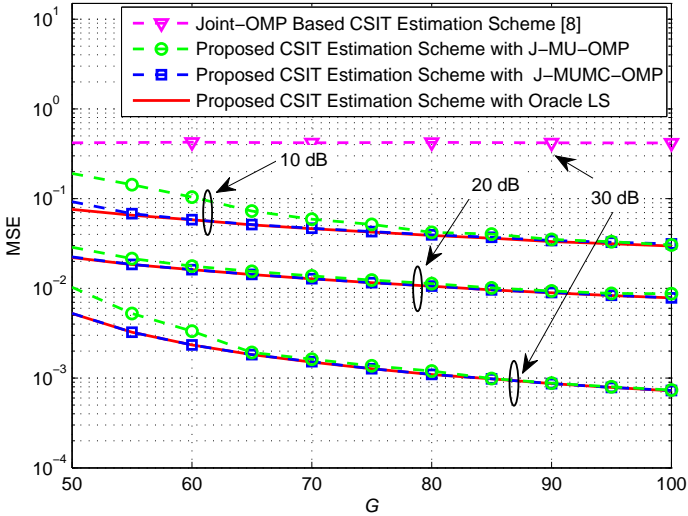}
\end{center}
%\vspace*{-5mm}
\caption{Comparison of channel estimation MSE performance of different CSIT estimation solutions versus $G$ at different $\rho_{\rm edge}$'s.}
\label{fig:mse_vs_T} % Fig 8
%\vspace*{-7mm}
\end{figure}
\subsection{Multi-Cell Joint Precoding}\label{S3.3}
In Section III-A, B, and C, we can use the low training overhead to estimate CSIT, which can be leveraged to perform multi-cell joint precoding to combat ICI.
Specifically, we consider: 1) each BS uses zero forcing (ZF) precoding to serve multiple users;
2) multiple users served by the BS using the same time-frequency resource should come from
different user groups to reduce the correlation of different users' channel vectors and enhance the system capacity;
3) each user is jointly served by multiple adjacent BSs according to the channel quality. The multi-cell joint precoding
can be integrated with the emerging cloud radio access network (C-RAN), where BS can be considered as the remote radio header (RRH) and a baseband unit (BBU) can be used to perform multi-cell joint precoding with centralized processing.
 %\vspace*{-2mm}
\section{Simulation Results}\label{S5}
 %\vspace*{-0mm}
In this section, we investigate the performance of the proposed
CS-based CSIT estimation scheme for downlink pilot decontamination in multi-cell FDD massive MIMO.
In simulations, we consider ${{L}}=7$ hexagonal cells, each BS
has $M=128$ antennas to simultaneously serve $N$ users, the carrier frequency is $f_c=2$~GHz,
the system bandwidth is $f_s=10$ MHz,
the maximum delay spread is $\tau_{\rm max}=5~\mu$s for typical urban scenario \cite{my_TSP}, $P=f_s\tau_{\rm max}=50$, the cell radius is 1 km, and the distance-based path loss between the $l$th BS and the $k$th user is ${\beta _{{\rm{PL}}}} = 1/({{{d^\alpha }}})$, where $d$ is the geographical distance between the $l$th BS and the $k$th user,
and the path loss exponent $\alpha$ is 3.8 dB/km.
Moreover, cell-edge users are considered in simulations since they suffer from the most severe downlink pilot contamination.
Specifically, we consider $K$ adjacent users as a group, and they are randomly distributed at the cell-edge of the central target BS with the geographical distance of 1 km.
For the proposed J-MUMC-OMP algorithm, ${\rho _{{\rm{th}}}}$ is set as 3, 5, 10, 10, and 10, ${\gamma _{{\rm{th}}}}$ is set as 0.006, 0.004, 0.002, 0.0004 and 0.0003 for $\rho_{\rm edge}=10~\rm{dB}$, $15~\rm{dB}$, $20~\rm{dB}$, $25~\rm{dB}$, and $30~\rm{dB}$, respectively, where $\rho_{\rm edge}$ is the cell-edge SNR associated with the central target BS. %Moreover, the %performance of the
%oracle LS estimator with the known support set of the angle-domain massive MIMO channels according to $\Pi$ is provided as the benchmark for comparison.
The joint-OMP based CSIT estimation scheme \cite{Rao1} only considering the single-cell scenario is provided for comparison. Besides, we also provide the so-called J-MU-OMP algorithm, which is a special case of the proposed J-MUMC-OMP algorithm when only the partially common sparsity among different users is considered.

 Fig.~\ref{fig:mse_vs_T} compares the channel estimation MSE performance of different CSIT estimation schemes,
 where $K=10$, $|{\Omega _{k,l}}|_c=6$, $\left\{ {{\Omega _{k,l}}} \right\}_{k = 1}^{K/2} = \Omega _l^1$, $\left\{ {{\Omega _{k,l}}} \right\}_{k = K/2+1}^{K} = \Omega _l^2$, and $|\Omega _l^1 \cap \Omega _l^2|_c=4$, $\forall l$ were considered. The oracle LS estimator with the known $\left\{ {{\Omega _{k,l}}} \right\}_{k = 1}^K$ for $l \in \Pi $ was adopted as the performance bound. From Fig.~\ref{fig:mse_vs_T}, it can be observed that the joint-OMP based CSIT scheme \cite{Rao1} suffers from downlink pilot contamination and works poorly. In contrast, J-MUMC-OMP and J-MU-OMP algorithms can effectively solve this issue thanks to the CS-based pilot design and CSIT estimation algorithm in multi-cell scenario. Especially, compared with J-MU-OMP algorithm, the proposed J-MUMC-OMP is
 capable of approaching the oracle LS performance bound when $G \ge 55$, since the common sparsity of angle-domain massive MIMO
 channels over different subcarriers is also leveraged for further improved performance.

\begin{figure}[!t]
\begin{center}
%\vspace*{-4mm}
\includegraphics[width=1\columnwidth, keepaspectratio]{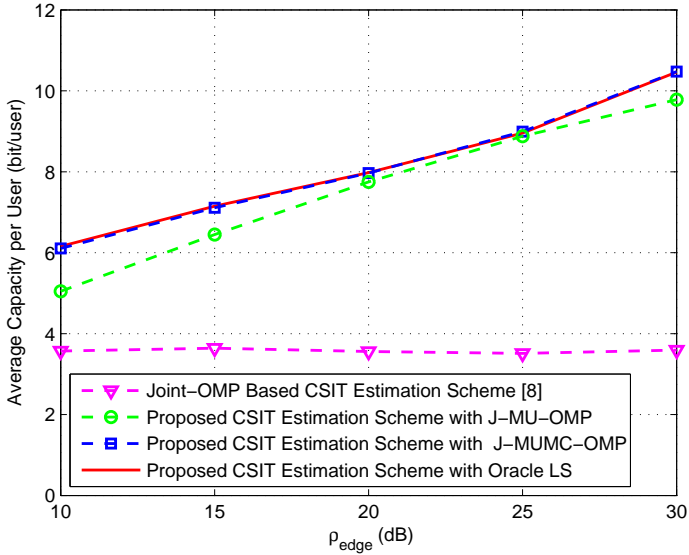}
\end{center}
%\vspace*{-5mm}
\caption{Comparison of downlink average throughput per user with multi-cell joint ZF precoding when $G=55$.}
\label{fig:ber_vs_snr} % Fig. 13
%\vspace*{-5mm}
\end{figure}

 Fig.~\ref{fig:ber_vs_snr} compares the downlink average throughput per user (bit/user) with different CSIT estimation schemes,
 where ZF precoding is used with $N=24$, and each user is jointly served by three best BSs according to their channel quality. It can be
 observed that the proposed J-MUMC-OMP based CSIT estimation scheme outperforms its counterparts,
 and its average throughput per user is capable of approaching that of the performance bound achieved by the oracle LS estimator.

%\vspace*{-2mm}
\section{Conclusions}\label{S6}

In this paper, we have proposed the CS-based CSIT estimation scheme for downlink pilot decontamination in multi-cell FDD massive MIMO systems, while existing schemes only consider the single-cell scenario and suffer from ICI.
We have exploited the common sparsity of angle-domain massive
MIMO channels over different
subcarriers and the partially common sparsity shared by adjacent
users.
By exploiting these sparsity features, we design the pilot signal and channel estimation algorithm under the CS framework. The proposed scheme
can reliably estimate multiple adjacent
BSs' channels for downlink pilot decontamination. Simulation results confirm that the
proposed solution outperforms existing schemes in multi-cell FDD
massive MIMO with low training overhead.

\section*{Acknowledgement}
The work was supported by the National Natural Science Foundation of China (NSFC) under Grants 62071044 and 61827901,
the Beijing Natural Science Foundation (BJNSF) under Grant L182024.

%\vspace*{-2mm}

%%\vspace*{-15mm}


\begin{thebibliography}{10}

%\bibitem{ }
%Z. Jiang, A. F. Molisch, G. Caire, and Z. Niu,
%``Achievable rates of FDD massive MIMO systems with spatial channel correlation,''
% {\em IEEE Trans. Wireless Commun.}, vol.~14, no.~5, pp.~2868--2882, May 2015.
\bibitem{Massve_MIMO_101} % 4
E. Bjornson, E. G. Larsson, and T. L. Marzetta,
 ``Massive MIMO: Ten myths and one critical question,'' {\em IEEE Commun. Mag.}, vol.~54, no.~2, pp.~114--123, Feb. 2016.

\bibitem{GFF_CM}
B. Wang et al.,
``Spatial-wideband effect in massive MIMO with application in mmWave systems,''
{\em IEEE Commun. Mag.}, vol. 56, no. 12, pp. 134--141, Dec. 2018.

%\bibitem{ } % 4
% H.~Yin, D.~Gesbert, M.~Filippou, and Y.~Liu, ``A coordinated approach to channel
% estimation in large-scale multiple-antenna systems,'' {\em IEEE J. Sel. Areas
% Commun.}, vol.~31, no.~2, pp.~264--273, Feb. 2013.
%
%\bibitem{FDDocloop} % 5
% J.~Choi, D.~J.~Love, and P.~Bidigare, ``Downlink training techniques for FDD
% massive MIMO systems: Open-loop and closed-loop training with memory,''
% {\em IEEE J. Sel. Topics Signal Process.}, vol.~8, no.~5, pp.~802--814, Oct. 2014.

\bibitem{MXS}
X. Ma et al.,
``Model-driven deep learning based channel estimation and feedback for millimeter-wave massive hybrid MIMO systems,''
{\em IEEE J. Sel. Areas Comm.}, doi: 10.1109/JSAC.2021.3087269.

\bibitem{FJ_TVT}
J. Fang, X. Li, H. Li, and F. Gao,
``Low-rank covariance-assisted downlink training and channel estimation for FDD massive MIMO systems,''
{\em IEEE Trans. Wireless Commun.}, vol. 16, no. 3, pp. 1935--1947, Mar. 2017.

%\bibitem{Shim}J. W. Choi, B. Shim, and Seok-Ho Chang,
% ``Downlink pilot reduction for massive MIMO systems via compressed sensing,''
% {\em IEEE Commun. Lett.}, vol.~19, no.~11, pp.~1889--1892, Nov. 2015.

\bibitem{Shen}W. Shen, L. Dai, Y. Shi, B. Shim, and Z. Wang,
 ``Joint channel training and feedback for FDD massive MIMO systems,''
 {\em IEEE Trans. Veh. Techn.}, vol. 65, no. 10, pp. 8762--8767, Oct. 2016.

\bibitem{WS_TVT}
S. Wu, H. Yao, C. Jiang, X. Chen, L. Kuang, and L. Hanzo,
``Downlink channel estimation for massive MIMO Systems relying on vector approximate message passing,"
{\em IEEE Trans. Veh. Techn.}, vol. 68, no. 5, pp. 5145--5148, May 2019.

\bibitem{WS_CL}
S. Hou, Y. Wang, T. Zeng, and S. Wu,
``Sparse channel estimation for spatial non-stationary massive MIMO channels,"
{\em IEEE Commun. Lett.}, vol. 24, no. 3, pp. 681--684, Mar. 2020.

%\bibitem{TSP_Masood}M. Masood, L. H. Afify, and T. Y. Al-Naffouri,
%``Efficient coordinated recovery of sparse channels in massive MIMO,''
%{\em IEEE Trans. Signal Process.}, vol. 63, no. 1, pp. 104--118, Jan. 2015.

\bibitem{Rao1} % 15
 X.~Rao and V.~K.~N.~Lau, ``Distributed compressive CSIT estimation and feedback for FDD
 multi-user massive MIMO systems,'' \emph{IEEE Trans. Signal Process.}, vol.~62, no.~12,
 pp.~3261--3271, Jun. 2014.

\bibitem{HY_Tcom}
Y. Han, J. Lee, and D. J. Love,
``Compressed sensing-aided downlink channel training for FDD massive MIMO systems,"
{\em IEEE IEEE Trans. Commun.}, vol. 65, no. 7, pp. 2852--2862, Jul. 2017.

\bibitem{Rao2} % 15
V.~K.~N.~Lau, S. Cai, and A. Liu, ``Closed-loop compressive CSIT estimation in
FDD massive MIMO systems with 1 bit
feedback,''\emph{IEEE Trans. Signal Process.}, vol.~64, no.~8,
 pp.~2146--2155, Apr. 2016.

\bibitem{my_TSP}Z. Gao, L. Dai, Z. Wang, and S. Chen,
``Spatially common sparsity based adaptive channel estimation and feedback for FDD massive MIMO,''
{\em IEEE Trans. Signal Process.}, vol. 63, no. 23, pp. 6169--6183, Dec. 2015.

\bibitem{GFF_TSP}
B. Wang, F. Gao, S. Jin, H. Lin, and G. Y. Li,
``Spatial- and frequency-wideband effects in millimeter-wave massive MIMO systems,''
{\em IEEE Trans. Signal Process.}, vol. 66, no. 13, pp. 3393--3406, Jul. 2018.
%\bibitem{myCL2}C. Qi, L. Wu, Y. Huang, and A. Nallanathan,
%``Joint design of pilot power and pilot pattern for sparse cognitive radio systems,''
%{\em IEEE Trans. Vehi. Techn.}, vol. 64, no. 11, pp. 5384-5390, Nov. 2015.

\bibitem{STR_CS} % 24
% M.~F.~Duarte and Y.~C.~Eldar, ``Structured compressed sensing: From theory to applications,''
% {\it IEEE Trans. Signal Process.}, vol.~59, no.~9, pp.~4053--4085, Sep. 2009.
M. Duarte, S. Sarvotham, D. Baron, M. Wakin, and R. Baraniuk,
``Distributed compressed sensing of jointly sparse signals,'' in  {\it Proc.
Asilomar Conf. Signals, Syst., Comput.}, 2005, pp. 1537--1541.


%\bibitem{SAMP} % 22
% T.~T.~Do, L.~Gan, N.~Nguyen, and T.~D.~Tran, ``Sparsity adaptive matching pursuit algorithm
% for practical compressed sensing,'' in {\em Proc. 42nd Asilomar Conf. Signal Syst. Comput.}
% (Pacific Grove, CA), Oct. 26-29, 2008, pp.~581--587.
%\bibitem{channel_model_for4g}L. Correia,
%{\it Mobile Broadband Multimedia Networks, Techniques, Models and Tools for 4G},
%San Diego, CA: Academic, 2006.



%\bibitem{cedu} % 25
% P.~Billingsley, {\em Probability and Measure}. Wiley, 1979.

%\bibitem{orthogonal} % 10
% H.~Minn and N.~Al-Dhahir, ``Optimal training signals for MIMO OFDM channel estimation,''
% {\em IEEE Trans. Wireless Commun.}, vol.~5, no.~5, pp.~1158--1168, May 2006.





\end{thebibliography}
\end{document}